\documentclass[a4paper, 12pt]{article}
\usepackage{ifthen}
\newboolean{pdflatex}
\setboolean{pdflatex}{true}
\newboolean{articletitles}
\setboolean{articletitles}{true}
\newboolean{uprightparticles}
\setboolean{uprightparticles}{false}
\newboolean{inbibliography}
\setboolean{inbibliography}{false}
\usepackage{cite}
\usepackage{mcite}
\usepackage{mciteplus}
\usepackage{geometry}
\usepackage[utf8]{inputenc}
\usepackage{graphics}
\usepackage{graphicx}
\usepackage{amsmath,amssymb}
\usepackage[normalem]{ulem}
\usepackage{dsfont}
\usepackage{hyperref}
\def\be{\begin{equation}}
\def\ee{\end{equation}}
\def\bea{\begin{eqnarray}}
\def\eea{\end{eqnarray}}

\def \bea{\begin{eqnarray}}
\def \beq{\begin{equation}}
\def \eea{\end{eqnarray}}
\def \eeq{\end{equation}}

\def \bma{\begin{matrix}}
\def \ema{\end{matrix}}
\def \({\left(}
\def \){\right)}
\def \[{\left[}
\def \]{\right]}

\def \be{\beta}

\def \Kp{K^+\pi^-}

\usepackage{xspace} 
\usepackage{upgreek}

\def\lhcb {\mbox{LHCb}\xspace}

\def\babar  {\mbox{BaBar}\xspace}
\def\belle  {\mbox{Belle}\xspace}
\def\besthree  {\mbox{BES-III}\xspace}
\def\belletwo  {\mbox{Belle-II}\xspace}

\def\cleoc   {\mbox{CLEO-c}\xspace}
\def\cdf    {\mbox{CDF}\xspace}

\def\lhc    {\mbox{LHC}\xspace}

\def\MagUp {\mbox{\em Mag\kern -0.05em Up}\xspace}

\ifthenelse{\boolean{uprightparticles}}%
{
 
 \def\Pgamma      {\ensuremath{\upgamma}\xspace}

 \def\Peta        {\ensuremath{\upeta}\xspace}

 \def\Ppi         {\ensuremath{\uppi}\xspace}

 \def\Pphi        {\ensuremath{\upphi}\xspace}                 
                  
 \def\Pchi        {\ensuremath{\upchi}\xspace}

 \def\PDelta      {\ensuremath{\Delta}\xspace}                 
 \def\PXi      {\ensuremath{\Xi}\xspace}                 
 \def\PLambda      {\ensuremath{\Lambda}\xspace}                 
 \def\PSigma      {\ensuremath{\Sigma}\xspace}                 
 \def\POmega      {\ensuremath{\Omega}\xspace}                 
 \def\PUpsilon      {\ensuremath{\Upsilon}\xspace}                 
 
 \def\PB      {\ensuremath{\mathrm{B}}\xspace}                 
                  
 \def\PD      {\ensuremath{\mathrm{D}}\xspace}

 \def\PK      {\ensuremath{\mathrm{K}}\xspace}

 \def\Pb      {\ensuremath{\mathrm{b}}\xspace}                 
 \def\Pc      {\ensuremath{\mathrm{c}}\xspace}                 
 \def\Pd      {\ensuremath{\mathrm{d}}\xspace}                 
 \def\Pe      {\ensuremath{\mathrm{e}}\xspace}

 \def\Ph      {\ensuremath{\mathrm{h}}\xspace}                 
 \def\Pi      {\ensuremath{\mathrm{i}}\xspace}

 \def\Pp      {\ensuremath{\mathrm{p}}\xspace}

 \def\Ps      {\ensuremath{\mathrm{s}}\xspace}                 
                  
 \def\Pu      {\ensuremath{\mathrm{u}}\xspace}

}
{
 
 \def\Pgamma      {\ensuremath{\gamma}\xspace}

 \def\Peta        {\ensuremath{\eta}\xspace}

 \def\Ppi         {\ensuremath{\pi}\xspace}

 \def\Pphi        {\ensuremath{\phi}\xspace}                 
                  
 \def\Pchi        {\ensuremath{\chi}\xspace}

 \mathchardef\PDelta="7101
 \mathchardef\PXi="7104
 \mathchardef\PLambda="7103
 \mathchardef\PSigma="7106
 \mathchardef\POmega="710A
 \mathchardef\PUpsilon="7107
                  
 \def\PB      {\ensuremath{B}\xspace}                 
                  
 \def\PD      {\ensuremath{D}\xspace}

 \def\PK      {\ensuremath{K}\xspace}

 \def\Pb      {\ensuremath{b}\xspace}                 
 \def\Pc      {\ensuremath{c}\xspace}                 
 \def\Pd      {\ensuremath{d}\xspace}                 
 \def\Pe      {\ensuremath{e}\xspace}

 \def\Ph      {\ensuremath{h}\xspace}                 
 \def\Pi      {\ensuremath{i}\xspace}

 \def\Pp      {\ensuremath{p}\xspace}

 \def\Ps      {\ensuremath{s}\xspace}                 
                  
 \def\Pu      {\ensuremath{u}\xspace}

}

\makeatletter
\ifcase \@ptsize \relax
  \newcommand{\miniscule}{\@setfontsize\miniscule{4}{5}}
\or
  \newcommand{\miniscule}{\@setfontsize\miniscule{5}{6}}
\or
  \newcommand{\miniscule}{\@setfontsize\miniscule{5}{6}}
\fi
\makeatother

\DeclareRobustCommand{\optbar}[1]{\shortstack{{\miniscule (\rule[.5ex]{1.25em}{.18mm})}
  \\ [-.7ex] $#1$}}

\def\en         {{\ensuremath{\Pe^-}}\xspace}   
\def\ep         {{\ensuremath{\Pe^+}}\xspace}

\def\g      {{\ensuremath{\Pgamma}}\xspace}

\def\uquark    {{\ensuremath{\Pu}}\xspace}
\def\uquarkbar {{\ensuremath{\overline \uquark}}\xspace}

\def\dquark    {{\ensuremath{\Pd}}\xspace}

\def\squark    {{\ensuremath{\Ps}}\xspace}

\def\cquark    {{\ensuremath{\Pc}}\xspace}
\def\cquarkbar {{\ensuremath{\overline \cquark}}\xspace}

\def\bquark    {{\ensuremath{\Pb}}\xspace}

\def\hadron {{\ensuremath{\Ph}}\xspace}
\def\hp {{\ensuremath{\hadron^{+}}}\xspace}
\def\hm {{\ensuremath{\hadron^{-}}}\xspace}

\def\pion   {{\ensuremath{\Ppi}}\xspace}
\def\piz    {{\ensuremath{\pion^0}}\xspace}

\def\pip    {{\ensuremath{\pion^+}}\xspace}
\def\pim    {{\ensuremath{\pion^-}}\xspace}

\def\kaon    {{\ensuremath{\PK}}\xspace}
  \def\Kbar    {{\kern 0.2em\overline{\kern -0.2em \PK}{}}\xspace}

\def\KorKbar    {\kern 0.18em\optbar{\kern -0.18em K}{}\xspace}

\def\Kp      {{\ensuremath{\kaon^+}}\xspace}
\def\Km      {{\ensuremath{\kaon^-}}\xspace}

\def\KS      {{\ensuremath{\kaon^0_{\mathrm{ \scriptscriptstyle S}}}}\xspace}

\def\Kstarm  {{\ensuremath{\kaon^{*-}}}\xspace}

\newcommand{\etaz}{\ensuremath{\Peta}\xspace}

  \def\Dbar    {{\kern 0.2em\overline{\kern -0.2em \PD}{}}\xspace}
\def\D       {{\ensuremath{\PD}}\xspace}

\def\DorDbar    {\kern 0.18em\optbar{\kern -0.18em D}{}\xspace}
\def\Dz      {{\ensuremath{\D^0}}\xspace}
\def\Dzb     {{\ensuremath{\Dbar{}^0}}\xspace}

\def\Dm      {{\ensuremath{\D^-}}\xspace}

\def\Dstarz  {{\ensuremath{\D^{*0}}}\xspace}

\def\Dsm     {{\ensuremath{\D^-_\squark}}\xspace}

\def\B       {{\ensuremath{\PB}}\xspace}
\def\Bbar    {{\ensuremath{\kern 0.18em\overline{\kern -0.18em \PB}{}}}\xspace}

\def\BorBbar    {\kern 0.18em\optbar{\kern -0.18em B}{}\xspace}

\def\Bu      {{\ensuremath{\B^+}}\xspace}

\def\Bp      {{\ensuremath{\Bu}}\xspace}

\def\Bd      {{\ensuremath{\B^0}}\xspace}
\def\Bs      {{\ensuremath{\B^0_\squark}}\xspace}

\def\Bds     {{\ensuremath{\B^0_{(\squark)}}}\xspace}

\def\chiczero {{\ensuremath{\Pchi_{\cquark 0}}}\xspace}

  \def\Y#1S{\ensuremath{\PUpsilon{(#1S)}}\xspace}

\def\FiveS {{\Y5S}}

\def\proton      {{\ensuremath{\Pp}}\xspace}
\def\antiproton  {{\ensuremath{\overline \proton}}\xspace}

\def\Xires       {{\ensuremath{\PXi}}\xspace}

\def\Lz          {{\ensuremath{\PLambda}}\xspace}
\def\Lbar        {{\ensuremath{\kern 0.1em\overline{\kern -0.1em\PLambda}}}\xspace}
\def\LorLbar    {\kern 0.18em\optbar{\kern -0.18em \PLambda}{}\xspace}

\def\Lb      {{\ensuremath{\Lz^0_\bquark}}\xspace}
\def\Lbbar   {{\ensuremath{\Lbar{}^0_\bquark}}\xspace}
\def\Lc      {{\ensuremath{\Lz^+_\cquark}}\xspace}

\def\Xibz    {{\ensuremath{\Xires^0_\bquark}}\xspace}

\newcommand{\decay}[2]{\ensuremath{#1\!\to #2}\xspace}         

\def\to                 {\ensuremath{\rightarrow}\xspace}

\def\order   {{\ensuremath{\mathcal{O}}}\xspace}

\def\CP                {{\ensuremath{C\!P}}\xspace}

\newcommand{\DGs}{{\ensuremath{\Delta\Gamma_{\squark}}}\xspace}

\def\AT#1     {\ensuremath{A_{\mathrm{T}}^{#1}}\xspace}           

\def\C#1      {\ensuremath{\mathcal{C}_{#1}}\xspace}                       
\def\Cp#1     {\ensuremath{\mathcal{C}_{#1}^{'}}\xspace}                    
\def\Ceff#1   {\ensuremath{\mathcal{C}_{#1}^{\mathrm{(eff)}}}\xspace}        
\def\Cpeff#1  {\ensuremath{\mathcal{C}_{#1}^{'\mathrm{(eff)}}}\xspace}       
\def\Ope#1    {\ensuremath{\mathcal{O}_{#1}}\xspace}                       
\def\Opep#1   {\ensuremath{\mathcal{O}_{#1}^{'}}\xspace}                    

\newcommand{\tev}{\ifthenelse{\boolean{inbibliography}}{\ensuremath{~T\kern -0.05em eV}}{\ensuremath{\mathrm{\,Te\kern -0.1em V}}}\xspace}
\newcommand{\gev}{\ensuremath{\mathrm{\,Ge\kern -0.1em V}}\xspace}
\newcommand{\mev}{\ensuremath{\mathrm{\,Me\kern -0.1em V}}\xspace}
\newcommand{\kev}{\ensuremath{\mathrm{\,ke\kern -0.1em V}}\xspace}
\newcommand{\ev}{\ensuremath{\mathrm{\,e\kern -0.1em V}}\xspace}
\newcommand{\gevc}{\ensuremath{{\mathrm{\,Ge\kern -0.1em V\!/}c}}\xspace}
\newcommand{\mevc}{\ensuremath{{\mathrm{\,Me\kern -0.1em V\!/}c}}\xspace}
\newcommand{\gevcc}{\ensuremath{{\mathrm{\,Ge\kern -0.1em V\!/}c^2}}\xspace}
\newcommand{\gevgevcccc}{\ensuremath{{\mathrm{\,Ge\kern -0.1em V^2\!/}c^4}}\xspace}
\newcommand{\mevcc}{\ensuremath{{\mathrm{\,Me\kern -0.1em V\!/}c^2}}\xspace}

\def\invfb   {\ensuremath{\mbox{\,fb}^{-1}}\xspace}

\def\invab   {\ensuremath{\mbox{\,ab}^{-1}}\xspace}

\def\order{{\ensuremath{\mathcal{O}}}\xspace}

\def\gsim{{~\raise.15em\hbox{$>$}\kern-.85em
          \lower.35em\hbox{$\sim$}~}\xspace}
\def\lsim{{~\raise.15em\hbox{$<$}\kern-.85em
          \lower.35em\hbox{$\sim$}~}\xspace}

\def\sqs   {\ensuremath{\protect\sqrt{s}}\xspace}

\def\degrees{\ensuremath{^{\circ}}\xspace}

\def\tell1  {TELL1\xspace}
\def\ukl1   {UKL1\xspace}

\begin{document}
\begin{titlepage}

\vspace*{-1.5cm}
\noindent
\begin{tabular*}{\linewidth}{lc@{\extracolsep{\fill}}r@{\extracolsep{0pt}}}
 & & CKM2018 \\  
 & & Summary of WG V \\
 & & \today \\ 
\hline
\end{tabular*}

\vspace*{4.0cm}

\begin{center}
{\Large\bf 
\boldmath Progress and challenges in the study of direct \CP violation and $\gamma$ determinations:  \\[2mm] 
summary of CKM 2018 working group V}
\end{center}

\vspace{0.5cm}
\begin{center}
K.~K.~Vos$^{1}$,
M.~Sevior$^{2}$,
S.~Perazzini$^{3,a}$
\bigskip\\
{\normalfont\itshape\footnotesize
$ ^{1}$University of Siegen, Germany\\
$ ^{2}$University of Melbourne, Australia\\
$ ^{3}$CERN, Switzerland\\
$ ^{a}$Sezione INFN di Bologna, Bologna, Italy\\
}
\end{center}

\vspace{0.8cm}
\begin{abstract}
\vspace{0.2cm}\noindent
Purely hadronic $B$ decays are vital to the study of \CP violation within and beyond the Standard Model of particle physics. One key contribution is the determination of of the CKM angle $\gamma$. In addition, very large direct \CP violation has been observed in regions of phase space of multi-body hadronic decays, the description of which remains a significant challenge. In this summary, we discuss the recent results and progress for the determination of $\gamma$ and direct CP violation in purely hadronic $B$ decays as presented at CKM 2018.  
 
\end{abstract}

\end{titlepage}

\newpage
\pagenumbering{arabic}

\section{Introduction}
Within the Standard Model of particle physics (SM), \CP violation is accommodated by the Cabibbo-Kobayashi-Maskawa (CKM) matrix. The violation of the \CP symmetry is a crucial element to explain the imbalance between matter and anti-matter in the universe, and its study forms an important part of the flavour physics program. The goal is to precisely determine the SM CKM parameters, and to obtain insights in possible new sources of \CP violation beyond the SM. 

Purely hadronic $B$ decays are vital to this endeavour. Decays like $B \to D K$, allow for a precise determination of the Unitary Triangle (UT) angle $\gamma$, which plays an important role in the study of \CP violation. At the moment, this angle is the least well known of the three UT angles, however due to significant experimental progress this is expected to change in the near future. 
From a theoretical point, the study of \CP violation in pure (charmless) hadronic $B$ decays is challenging due to the strong decay dynamics of these decays.   
The incredible amount of data available from LHCb and the B-factories provides a perfect set-up, but requires reliable and precise theoretical predictions and poses experimental challenges. In this summary, we present recent experimental and theoretical progress, and give a brief outlook for the exciting future to come.



\section{Determination of $\gamma$}\label{sec:gammadet}
One of the key parameters of the CKM matrix is the Unitarity Triangle (UT) angle $\gamma$: 
\begin{equation}
\gamma = \rm{arg} \left(-\frac{V_{ud} V_{ub}^*}{V_{cd}V_{cb}^*}\right) \ .
\end{equation}
It can be determined in a theoretically clean way from tree-level \decay{\B}{\D^{(*)}\kaon^{(*)}} and \decay{\B}{\D^{(*)}\pion} decays using the interference between the \decay{\bquark}{\uquark\cquarkbar\squark(\dquark)} and \decay{\bquark}{\cquark\uquarkbar\squark(\dquark)} transitions. The theoretical uncertainty on $\gamma$ determined from these decays is expected to be $\delta\gamma/\gamma\lesssim 10^{-7}$ as no penguin operators contribute to the decay amplitudes, and in addition electroweak box corrections are shown to be tiny \cite{Bro13, Brod:2014qwa}. However, it has been pointed out that new physics contributions to the tree-level Wilson coefficients $C_1$ and $C_2$ might actually influence the determination of $\gamma$ from tree decays \cite{BLTXW, Gilberto}.  

Several methods to determine $\gamma$ have been proposed \cite{GLW1, GLW2, Gir03, ADS1, *ADS2},  differing in the type of \D decays that are probed: the GLW \cite{GLW1, GLW2} method, exploiting \D decays to \CP-eigenstates; the ADS\cite{ADS1, *ADS2}, using doubly Cabibbo-suppressed decays; the GGSZ\cite{Gir03},  which exploits the three-body \D decays to self-conjugate modes, such as \decay{\Dz}{\KS\hp\hm} (\hadron=\kaon, \pion). The best sensitivity is achieved by combining the various methods and decay modes as each of them has a different sensitivity to $\gamma$ depending on the \CP-conserving parameters entering the decay amplitudes. The latest combination of $\gamma$ measurements by the LHCb collaboration~\cite{LHCb-CONF-2018-002} yield
\begin{equation}\label{eq:lhcbgamma}
    \gamma = (74.0^{+5.0}_{-5.8})^{\circ},
\end{equation}
which is in very good agreement with the world averages $\gamma = (71.1^{+4.6}_{-5.3})^{\circ}$~\cite{HFLAV16}, $\gamma = (73.5^{+4.2}_{-5.1})^{\circ}$~\cite{CKMfitter2015} and $\gamma = (70.0 \pm 4.2)^{\circ}$~\cite{UTfit-UT}. The updated \lhcb combination includes also three recent LHCb analyses of \decay{\Bp}{D^{(*)}K^{(*)\pm}} decays~\cite{Aaij:2018uns,Aaij:2017glf,Aaij:2017ryw}, all based on data from Run1 and Run2, and corresponding to an integrated luminosity of \proton\proton collisions of 1\invfb at $\sqs=7\tev$, $2\invfb$ at $\sqs=8\tev$, and $2\invfb$ at $\sqs=13\tev$. Particularly interesting is the analysis reported in Ref.~\cite{Aaij:2017ryw}, where the \decay{\Dstarz}{\Dz\piz} and \decay{\Dstarz}{\Dz\gamma} decays are exploited without reconstructing the \piz or the \g. In fact, the invariant-mass distribution of \decay{\Bp}{\Dstarz\hp} decays with a partially reconstructed \Dstarz has a peculiar shape, which allow for the separation of the \decay{\Bp}{\Dstarz\hp} from backgrounds and fully reconstructed \decay{\Bp}{\Dz\hp} decays. With the full LHCb Run 2 data, the attainable uncertainty on $\gamma$ from this single decay is expected to be between 3\degrees and 4\degrees.

At the moment, the $\gamma$ combination from \lhcb in Eq.~\eqref{eq:lhcbgamma} is dominated by the results of GLW, ADS and GGSZ analyses, however, also other \decay{\B}{\D\hadron} modes contribute. Of those modes, very interesting are the tagged time-dependent analyses of the \decay{\Bs}{\Dsm\Kp}~\cite{Aaij:2017lff} and \decay{\Bd}{\Dm\pip}~\cite{Aaij:2018kpq} decays. These analyses provide valuable information on both $\gamma$ and $-2\beta_s$ (for the \Bs mode), and on both $\gamma$ and $\beta$ (for the \Bd mode) with minimal external inputs~\cite{Fleischer:2003yb,Dunietz:1987bv,Aleksan:1991nh,DeBruyn:2012jp}. In particular, the \Bs mode, which can uniquely be probed at \lhcb, allows for a determination of $\gamma$ with a precision of approximately 20\degrees with the current experimental precision using as the only external input the value $-2\beta_s$. Projections show that a precision of 2.5\degrees on $\gamma$ is achievable at the end of Run3 of \lhcb from this channel alone~\cite{LHCb-U2}.

With the increasing precision of the collected data sets, it will be possible to exploit also new decay modes to determine $\gamma$. Nice examples are \Bd and \Bs decays to \Dzb\pip\pim and \Dzb\Kp\Km final states. A time-dependent analysis of the Dalitz plane of these decays provides significant information on $\gamma$, $\beta$ and $-2\beta_s$~\cite{Nandi:2011uw,GLW2}. The first observation of the \decay{\Bs}{\Dzb\Kp\Km} decay is reported by \lhcb analyzing the full Run1 sample~\cite{Aaij:2018rol}. Restricting to the \Kp\Km mass region around the $\Pphi(1020)$ also the \decay{\Bs}{\Dstarz\Kp\Km} decay is observed for the first time\footnote{The technique of partially reconstructing \decay{\Dstarz}{\Dz\piz} and \decay{\Dstarz}{\Dz\g} decays is used.}. Even though the statistics are still insufficient to fully exploit the full potential of these decays, the long term prospects are encouraging given the large sensitivity to $\gamma$ provided by these modes~\cite{GLW2,LHCb-U2}.

\begin{figure}[t]
  \begin{center}
    \includegraphics[width=0.5\textwidth]{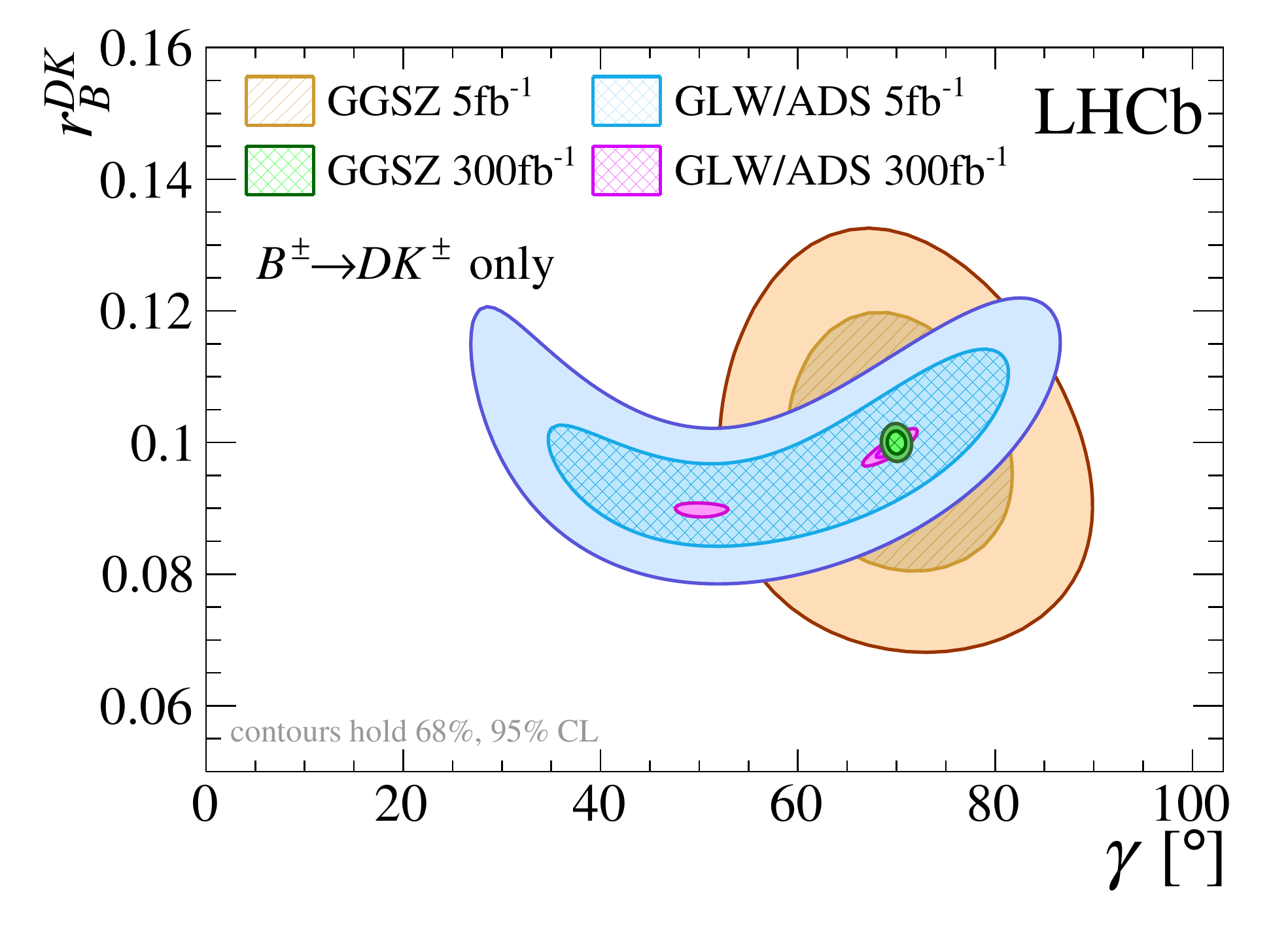}
 \end{center}
  \caption{\small Comparison between the current results of \lhcb using the GGSZ and the GLW/ADS methods, alongside with their future projections with 300~\invfb, in the plane of $\gamma$ vs. $r^{\D\kaon}_{\B}$. Plot taken from ~{\protect\cite{LHCb-U2}}. Note: the parameter $r_{\B}^{\D\kaon}$ is an additional hadronic parameter quantifying the ratio between the amplitudes of the suppressed \decay{\bquark}{\uquark} and favoured \decay{\bquark}{\cquark} transitions of the \B decay.}
  \label{fig:gammacombo}
\end{figure}

New analyses of \decay{\B}{\D\hadron} decays are also available from the \belle experiment~\cite{Resmi:2018mym}. The much cleaner environment of \ep-\en collisions and the fixed centre-of-mass energy allow to work with \D-meson decays that involve \g and \piz in the final state, that are very challenging for \lhcb, like \decay{\Dz}{\Km\pip\piz} or \decay{\Dz}{\KS\pip\pim\piz}. This last final state is very interesting given its relatively large branching fraction of 5.2\%. A precision on $\gamma$ of 25\degrees is expected from this channel alone with the full \belle data set, while a 4.4\degrees precision is foreseen if projecting to the 50\invab of integrated luminosity of \belletwo~\cite{Resmi:2017qxf}. Including all the main \decay{\B}{\D\hadron} modes, \belletwo is expected to measure the angle $\gamma$ with a precision of 1\degrees with 50\invab of integrated luminosity~\cite{Kou:2018nap}. The projection for \belletwo are encouragingly supported by the first run of data taking. With an integrated luminosity of approximately 0.5~\invfb \B- and \D-meson decays have been rediscovered. A promising yield of 254 \decay{\B}{\D\pion} candidates is observed with the \D mesons reconstructed using various final states.

\begin{figure}[t]
  \begin{center}
    \includegraphics[width=0.5\textwidth]{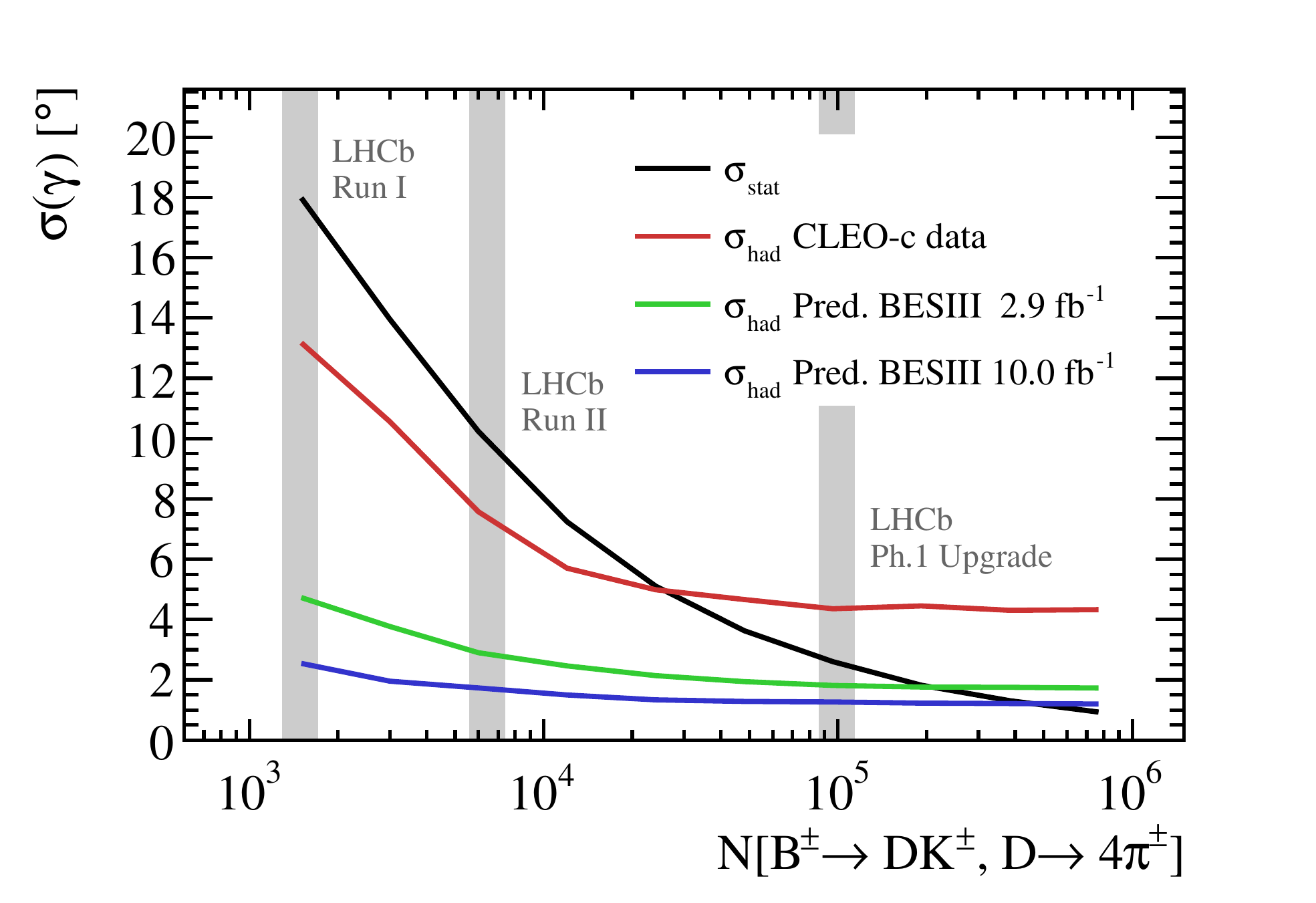}
 \end{center}
  \caption{\small Contribution of \cleoc and \besthree to the determination of $\gamma$ from \decay{\Bp}{\Dz\Kp} decay with the \Dz meson going to \pip\pim\pip\pim final state. The uncertainty on $\gamma$ as a function of the signal yields is determined considering different contribution from the charm factories. Plot taken from ~{\protect\cite{Harnew:2017tlp}}.}
  \label{fig:gammabes}
\end{figure}

As already mentioned, thanks to impressive data set that will be collected by \belletwo and the \lhcb upgrade, the determination of the angle $\gamma$ will reach an impressive precision of about 1\degrees for each experiment in the next few years. Even more impressive are the prospects for a second upgrade of \lhcb; collecting an integrated luminosity of 300\invfb would shrink the uncertainty on $\gamma$ to 0.35\degrees~\cite{LHCb-U2}. Figure~\ref{fig:gammacombo} shows the comparison between the current determination of $\gamma$ from \lhcb and the expected determination with 300~\invfb of integrated luminosity. However, the analyses that mainly contribute to the $\gamma$ determination require external inputs, that will soon become the limiting factor to achieve such a precision. The best way to determine these input parameters is to study quantum-correlated charm decays, that can be produced at \ep-\en colliders operating at the $\psi(3770)$ resonance threshold. Decays of $\psi(3770)$ to two neutral charm mesons allow the measurement of the strong phases, the coherence factors and the \CP content of multibody charm decays. In this respect the r\^{o}le of \besthree, and its predecessor \cleoc, will be crucial. The current contribution of \cleoc is quantified in an uncertainty of about 2\degrees on the overall error on $\gamma$~\cite{Malde:2223391}. Fully exploiting the full sample collected by \besthree, corresponding to about 10 times the integrated luminosity of \cleoc, will allow to reach a sub-degree precision on $\gamma$. As an example, in Figure~\ref{fig:gammabes} the contribution of \besthree and \cleoc to the determination of $\gamma$ from the \decay{\Bp}{\Dz\Kp} decay with the \Dz meson going to \pip\pim\pip\pim final state, is shown~\cite{Harnew:2017tlp}.

\section{\boldmath{\CP} violation in two-body $B$ decays}
Two-body \B decays are theoretically challenging to describe as both the initial and final states are purely hadronic. These decays have been studied in various approaches such as QCD factorization (QCDF)~\cite{Beneke:1999br,Beneke:2000ry,Beneke:2003zv}, pQCD~\cite{Keum:2000ph,Keum:2000wi,Lu:2000em} or by using flavour symmetries of the light quarks~\cite{Zeppenfeld:1980ex}. Here we focus on the status and prospects for QCD factorization \cite{Tobias} and recent developments for $B\to \pi K$ decays using flavour symmetries \cite{Fleischer:2018jne}.  

QCD factorization makes use of the heavy $B$ meson mass to make a systematic expansion both in $\Lambda/m_b$ as well as in $\alpha_s$. Importantly, at leading order in $\Lambda/m_b$ the strong phases required to generate \CP violation are only generated at order $\alpha_s$, making the study of these contributions crucial. While the topological tree amplitudes are known already at NNLO level ~\cite{Bell:2007tv,Bell:2009nk,Bell:2009fm,Beneke:2009ek,Beneke:2005vv,Kivel:2006xc,Beneke:2006mk,Jain:2007dy,Pilipp:2007mg,Kim:2011jm}, the leading penguin amplitudes represent the last missing ingredient to establish QCD factorization at the NNLO level to leading power in $\Lambda_{\rm QCD}/m_b$. The calculation of the vertex correction to the latter amounts to a genuine two-loop, two-scale problem including a kinematic threshold, and applies state-of-the-art multi-loop techniques. Recently, also the leading penguin amplitudes were calculated ~\cite{Bell:2014zya,Bell:2015koa}. The NNLO correction turns out to be sizable, but there is no breakdown of the perturbative expansion \cite{Bell:2015koa}. The sizable NNLO effects in the amplitude gets diluted in quantities like amplitude ratios, direct \CP asymmetries and branching ratios.

The key question is how to improve QCDF-based predictions for non-leptonic B-decays in the future \cite{Tobias}. On the perturbative side, the NNLO QCD correction to the power-suppressed but chirally enhanced (and therefore phenomenologically relevant) penguin amplitude $a_6$ are still missing, as well as NLO QED corrections. Moreover, it could be beneficial to combine insights from two-body decays with the QCDF formulation of three-body decays~\cite{Krankl:2015fha,Klein:2017xti}. The most urgent issue in two-body charmless nonleptonic B-decays in QCDF is, however, improving our poor understanding of power corrections. To date their inclusion amounts to a rather crude parametrisation~\cite{Beneke:2001ev} or to a data-driven approach~\cite{Bobeth:2014rra}. An understanding of power corrections in QCDF on field-theoretic grounds is therefore highly desired, and ideas in this direction exist based on the `collinear anomaly'~\cite{Becher:2010tm,Chiu:2012ir} in collider physics. The long-term goal is to obtain a comprehensive QCDF analysis of all channels and observables with a satisfactory treatment of power corrections. 

In this respect, phenomenological studies using flavour symmetries can provide valuable insights into the strong decay dynamics of pure hadronic decays and non-perturbative contributions. Specifically interesting are $B \to \pi K$ decays, which are dominated by QCD penguin topologies due to a CKM suppression of the tree topologies, and which were studied extensively~\cite{Fleischer:2017vrb,Fleischer:2018bld,Gronau:1994bn,Fleischer:1995cg,Neubert:1998pt,Neubert:1998jq,Buras:1998rb,Beneke:2001ev,Buras:2003dj,Buras:2004ub,Gronau:2005kz,Gronau:2006xu,Fleischer:2008wb}. In fact, using an isospin relation between the neutral $B\to \pi K$ decays, a correlation between the direct and mixing-induced \CP asymmetries of \decay{\Bd}{\piz\KS} is found, which is in tension with the experimental data \cite{Fleischer:2008wb, Fleischer:2017vrb,Fleischer:2018bld}. Interestingly, this tension has grown stronger over the years in particular due to more precise determination of $\gamma$, which emphasizes the importance of improved $\gamma$ determinations for phenomenological studies. The tension cannot trivially be resolved by a change in the data, as was recently discussed \cite{Fleischer:2017vrb, Fleischer:2018bld, Fleischer:2018jne}. Here a state-of-the-art analysis of $B\to \pi K$ using flavour SU(3) symmetry was performed, including the electroweak penguin (EWP) contributions that are at the same level as the tree contribution in these decays. In addition, the angle $\phi_\pm$ between the amplitude $A(B_d^0 \to \pi^- K^+)$ and its \CP-conjugate was studied and found to provide an additional stringent constraint which increases the tension between the SM prediction and the current experimental data. An intriguing possibility is that this discrepancy is signalling New Physics, for which a modified EWP sector is a prime candidate. 

The $B\to \pi K$ decays, therefore, provide an excellent opportunity to study the parameters that govern these EWP contributions. Isospin relations between the $B\to\pi K$ amplitudes can be used to determine them in a theoretically clean way, using only minimal SU(3) assumptions \cite{Fleischer:2008wb, Fleischer:2017vrb,Fleischer:2018bld}. In this new strategy, the mixing-induced \CP asymmetry of $B_d^0 \to \pi^0 K_{\rm S}$ plays an important role, offering exciting prospects for Belle II. Although in general, theoretical uncertainties are difficult to control in non-leptonic $B$ decays, in this new method, the theory uncertainty is expected to be at the same level as the experimental precision that can be reached at the end of Belle II \cite{Abe:2010gxa}. The new strategy offers exciting possibilities for the nearby future and would potentially allow for the observation of new \CP-violating physics in the $B\to \pi K$ system.

From the experimental point of view, charmless non-leptonic \B decays are also challenging. Main experimental issues are related to the cross-contamination of the different final states that need to be suppressed with excellent particle identification capabilities. This is complicated by the fact that the typical branching ratios of these decays vary by about three order to magnitudes (between $10^{-5}$ and $10^{-8}$) making some of the decay modes the main source of background to other channels. New results have been shown by \lhcb and \belle, plus interesting projections on the capabilities of \belletwo. Updated measurement of \CP violation in \Bds meson decays to \pip\pim, \Kp\Km and \Kp\pim final states have been reported by \lhcb~\cite{Aaij:2018tfw}. This includes the measurement of the time-integrated \CP asymmetries of the \decay{\Bd}{\Kp\pim} and \decay{\Bs}{\pip\Km} decays, and the time-dependent \CP asymmetries of the \decay{\Bd}{\pip\pim} and \decay{\Bs}{\Kp\Km} decays. All the measurements are world's best and the time-integrated \CP asymmetries of the \decay{\Bd}{\Kp\pim} and \decay{\Bs}{\pip\Km} decays are also leading the world averages of these quantities. They are used to perform the test of the Standard Model suggested in Ref.~\cite{Lipkin:2005pb}, finding no evidence of deviations from the Standard Model. The measurement of the \CP violation parameters in the \decay{\Bs}{\Kp\Km} decay is unique to \lhcb and constitutes the strongest evidence of time-dependent \CP violation in the \Bs-meson sector to date, with a significance of 4 standard deviations. These new measurements will enable improved constraints to be set on the CKM \CP-violating phases, using processes whose amplitudes receive significant contributions from loop diagrams both in the mixing and decay of \Bds mesons~\cite{Fleischer:2010ib,Ciuchini:2012gd,Aaij:2014xba}. 

\begin{figure}[t]
  \begin{center}
    \includegraphics[width=0.5\textwidth]{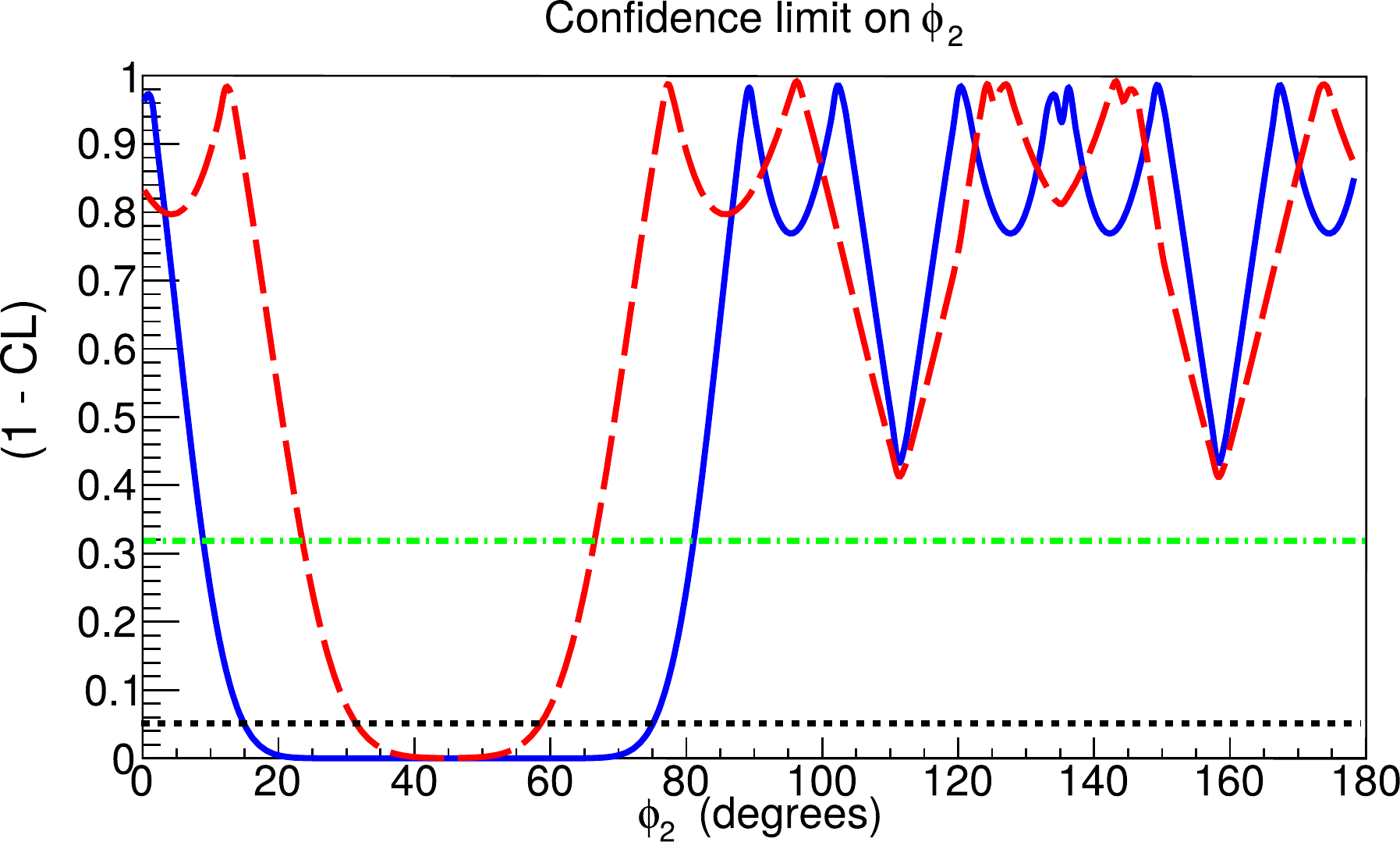}
 \end{center}
  \caption{\small Scan of the confidence level for $\alpha/\phi_2$ using only data
from \decay{\B}{\pion\pion} measurements of the \belle experiment. The dashed red curve shows the previous constraint, while the solid blue curve includes the newest measurement of the \decay{\Bd}{\piz\piz} decay. Plot taken from ~{\protect\cite{Julius:2017jso}}.}
  \label{fig:alphapipi}
\end{figure}

As already mentioned in Section~\ref{sec:gammadet} the much cleaner environment of \ep-\en colliders allows \belle to perform analyses with neutral particles in the final states, that would be very challenging for \lhcb, like \etaz\piz, \etaz\etaz and \piz\piz. Particularly interesting is the analysis of the \decay{\Bd}{\piz\piz} decay, whose \CP asymmetry and branching fraction are relevant input to the determination of the CKM angle $\alpha$~\cite{Gronau:1990ka} and to the analysis already mentioned in Ref.~\cite{Ciuchini:2012gd,Aaij:2014xba}. The updated result from \belle on the branching ratio and the direct \CP asymmetry of this decay~\cite{Julius:2017jso} provide a relevant contribution to the determination of $\alpha$, as visualised in Figure~\ref{fig:alphapipi}. However, the real breakthrough is expected with \belletwo, when the much larger statistics will allow to exploit the Dalitz decay \decay{\piz}{\g\ep\en}. This makes it possible to perform a time-dependent analysis of the \decay{\Bd}{\piz\piz} decay, measuring both the direct and mixing-induced \CP asymmetries. Finally, a precision on $\alpha$ of about 2\degrees is expected using only the \decay{\B}{\pion\pion} system, while the overall precision on $\alpha$ is expected to be below 1\degrees~\cite{Kou:2018nap}.

Another interesting channel that is challenging for \lhcb, given the reduced acceptance of the experiment to the long living \KS, is the \decay{\Bs}{\KS\KS} decay. This decay proceeds through pure penguin topologies and thus the measurement of its branching fraction and \CP asymmetries has the potential to probe several scenarios of physics beyond the Standard Model~\cite{Chang:2013hba,Baek:2006pb,Hayakawa:2013dxa}. The \belle experiment, using the 121.4\invfb of luminosity collected at the \FiveS resonance, reported the first observation of this decay with a significance of more than 5 standard deviations~\cite{Pal:2015ghq}. Even though the resolution on the decay time achievable by \belle (and \belletwo) is not sufficient to resolve the fast oscillation of the \Bs meson, a measurement of its effective lifetime is possible. Thanks to the non-zero decay-width difference \DGs of the \Bs meson, measuring the effective lifetime allows to extract information also on \CP violation. Approximately 1000 \decay{\Bs}{\KS\KS} signal candidates are expected with the full \belletwo sample.


\section{\boldmath{\CP}  violation in three-body $B$ decays}
Multibody decays form a large part of the non-leptonic $B$-meson branching fraction. Especially, in the study of \CP violation they are interesting because due to their non-trivial kinematics, these decays contain more information on strong phases than two-body decays. For charmless $B\to hhh$ decays, interesting and rich patterns of direct \CP violation, ($A_{CP}$), across the Dalitz distribution were indeed observed \cite{LHCbCPV2014}. In particular, a clear correlation  
between the channels $B^\pm \to K^\pm \pi^+\pi^-$ and  $B^\pm \to K^\pm K^+ K^-$ decays was observed, where the CP asymmetries carry opposite signs. This is a signature of the rescattering process  
$\pi^+\pi^-\to K^+K^-$ and a consequence of the $CPT$ symmetry, which requires the equal total decay rates between particles and antiparticles \cite{BOT}. Theoretically, such three-body decays are challenging to describe and a large variety of approaches are considered (see e.g. \cite{Uspin3, BOT, Boito:2017jav}). Recently, three-body decays were also studied for the first time in a QCD factorization approach \cite{Krankl:2015fha,Klein:2017xti}, which requires new non-perturbative objects as the $B\to hh$ and $2h$ distribution amplitudes.

To fully understand the different CP patterns and distribution of the strong phases across the Dalitz plot a full amplitude analysis is required. For $B^{\pm} \to \pi^-\pi^+\pi^{\pm}$ and $B^{\pm} \to K^-K^+\pi^{\pm}$, new analyses by the LHCb Collaboration are expected to appear soon \cite{Alberto}. Of particular note are the observation of very large direct CP-violation, (approaching 1), in regions of phase space for the $B^{\pm} \to K^{+}K^{-} \pi^{\pm}$ and $B^{\pm} \to \pi^{+} \pi^{-} \pi^{\pm}$ modes \cite{Alberto} \cite{Hsu:2017}. The interpretation of these decays is challenging, given the lack of proper description for all the components (in particular scalar resonances) contributing to the Dalitz plot distributions. The case of the $B^{\pm} \to K^{+}K^{-} \pi^{\pm}$ decay, is particularly challenging as the yield shows an order of magnitude increase above phase space below $M_{KK} < 1.1 \text{GeV}$ as well as $A_{CP} = -0.90\pm 0.17 \pm 0.04$ \cite{Hsu:2017}. Thus two amplitudes are required to provide an order of magnitude increase  below $M_{KK} < 1.1 \text{GeV}$. One where the weak phase phase changes sign between particles and anti-particles and the other where it does not.

The analysis of $B^{\pm} \to K^-K^+K^{\pm}$ is ongoing. In this decay, the rescattering of charm penguins might play an important role in the generation of strong phases \cite{Patricia, charmScat}. In fact, the CP asymmetry of $B^{\pm} \to K^-K^+K^{\pm}$ changes sign close to the $D\bar{D}$ open threshold around an invariant $KK$-mass of $4$ GeV \cite{LHCbCPV2014}. Therefore, at these high energies $D\bar{D} \to K\bar{K} (\pi^-\pi^+)$ may be important, in analogy to $\pi\pi - KK$ rescattering at low energies. The effect of this $D\bar{D}$ rescattering, can be parameterized by a pure hadronic triangle loop ~\cite{charmScat}, which would generate a distinct structure in the Dalitz plot. In particular, the phase of the hadronic amplitude changes sign across the $D\bar{D}$ threshold. This triangle parameterization could be confirmed by data, using amplitude analysis of the  $B^{\pm} \to K^-K^+K^{\pm}$ channel. Interestingly, a similar contribution would also influence the rare $B_c \to K^-K^+\pi^+$ decay \cite{CharmBc}. New data from LHCb, would be able to test the charm rescattering parametrization. 
 
\subsection{Extracting $\gamma$ from three-body decays }

The conventional methods to extract $\gamma$ from hadronic \decay{\B}{\D\hadron} decays are theoretically clean because these decays proceed through pure tree-level transitions. Besides, using flavour symmetries, it is also possible to extract $\gamma$ from penguin dominated charmless two-body \B decays \cite{Uspin2,Fleischer:2016ofb}, but also from three-body \B decays~\cite{Uspin3, Bhattacharya:2018pee}. Because of the relevant contributions of penguin topologies to the decay amplitudes, these latter decays might be sensitive to new particles that could enter the loops as virtual contributions, making very interesting to compare extractions of $\gamma$ from tree- and loop-dominated decays. 

The angle $\gamma$ can be extracted from \decay{\B}{\hadron\hadron\hadron} decays using a diagrammatic analysis, and by relating the hadronic parameters of the different decays by SU(3) flavour symmetry~\cite{Uspin3, Bhattacharya:2018pee, 3bgrp}. In this way, by using the available branching ratios, direct and mixing-induced \CP asymmetries, $\gamma$ can be determined from data~\cite{SU3ga}. Thanks to the number of observables that over-constrain the system of equations, it is also possible to introduce a parameter that takes into account possible SU(3)-breaking effects, hence relaxing the theoretical assumptions of the method. An implementation of this method using \babar data for \decay{\Bp}{\Kp\pip\pim}, \decay{\Bd}{\KS\pip\pim}, \decay{\Bd}{\Kp\pim\piz}, \decay{\Bd}{\KS\KS\KS} and \decay{\Bd}{\KS\Kp\Km}~\cite{Babar1,*Babar2,*Babar3,*Babar4} was first presented in Ref.~\cite{SU3ga, Bhattacharya:2018pee}. In the SU(3) limit, the momentum dependent hadronic parameters are equal in each point in the Dalitz plot, making it possible to extract $\gamma$ (and the SU(3)-breaking parameter) independently in different regions. The key advantage is that the extracted values of $\gamma$ can then be averaged over the entire Dalitz plot, and thereby increasing the precision. 

Recently, a new analysis using this method was performed, which takes into account several systematic uncertainties~\cite{Emilie, Bertholet:2018tmx}. By combining several hundred sets of points on the Dalitz plane, and taking into account the correlations among the different points, six possible solutions for $\gamma$ with a $\order(10\degrees)$ precision were found. The uncertainty is comparable with measurements of $\gamma$ from decays including loops processes and allows for a comparison with the world average from measurements obtained with tree-dominated decays~\cite{CKMfitter2015,UTfit-UT}. Very interesting is the possibility of the analysis to quantify the strength of SU(3)-breaking effects. Although locally SU(3) breaking might be large, it was found that averaging over a large number of points in the Dalitz plane, the breaking effects are $\order(6\%)$~\cite{Emilie,Bertholet:2018tmx}. 

Efforts to included anti-symmetric states or mixed states are undertaken, which may help to decrease the statistical uncertainties and reduced the number of solutions. It would be interesting to perform this analysis using LHCb and Belle-II data by extracting $\gamma$ directly from a simultaneous fit.

\subsection{Experimental status}

\begin{figure}[t]
  \begin{center}
    \includegraphics[width=0.35\textwidth]{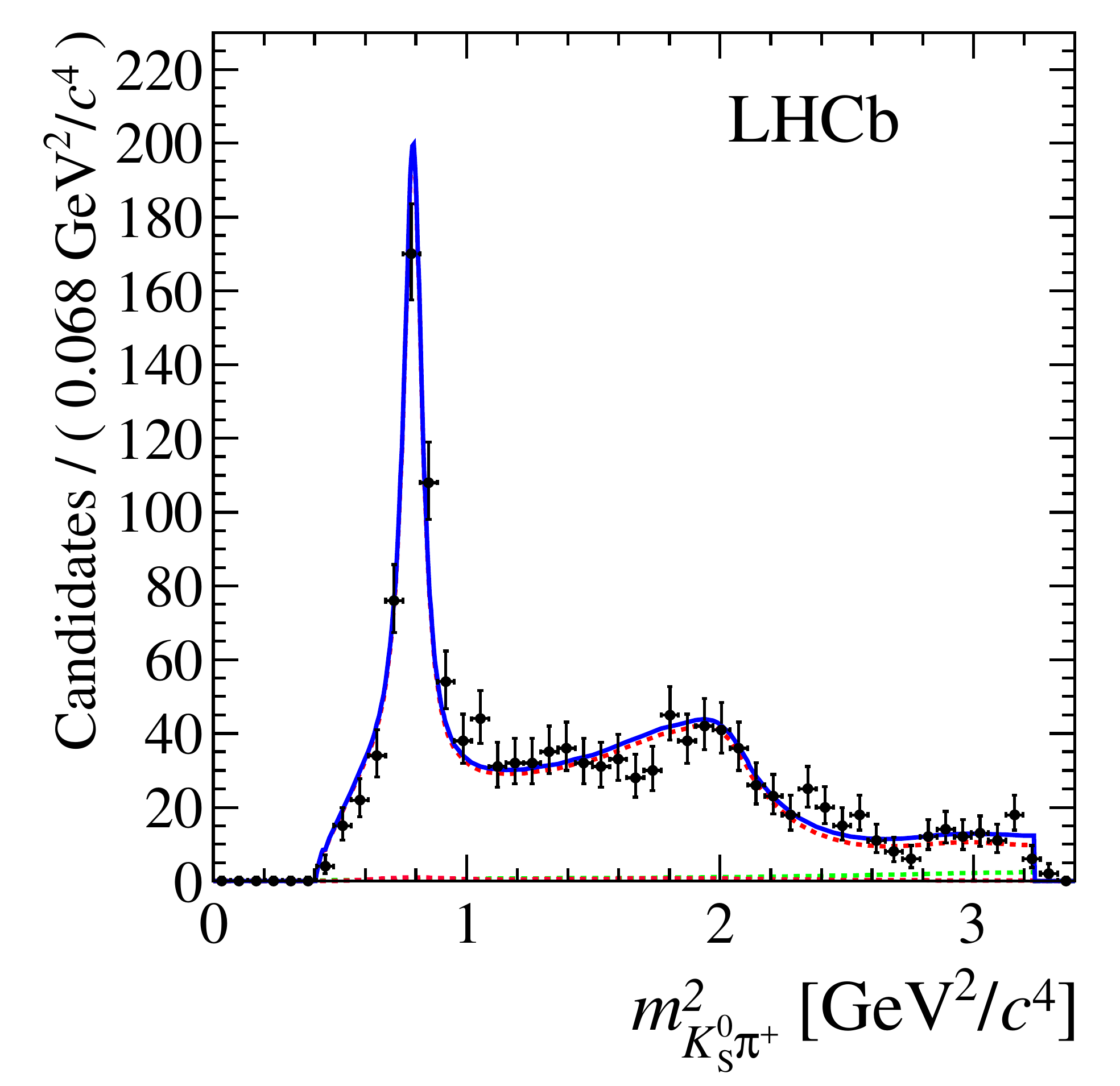}
    \includegraphics[width=0.35\textwidth]{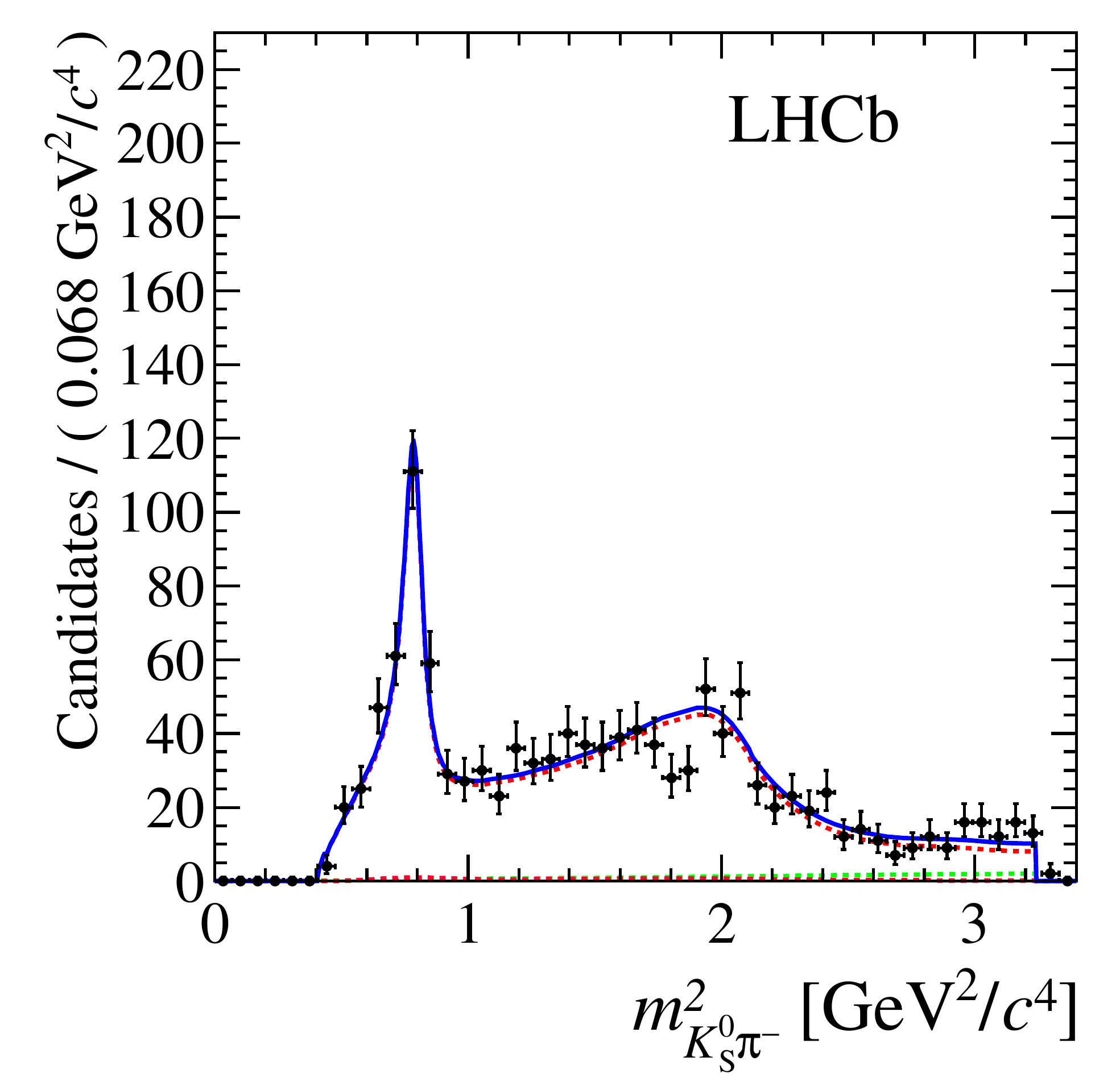}
 \end{center}
  \caption{\small Projections of the \decay{\Bd}{\KS\pip\pim} data points and best fit result (blue line) onto the Dalitz-plot variables (left) $m^2_{\KS\pip}$ and (right) $m^2_{\KS\pim}$. Projections are restricted to the two-body low invariant-mass regions. Plot taken from ~{\protect\cite{Aaij:2017ngy}}.}
  \label{fig:kspipiCPV}
\end{figure}

From the experimental point of view the study of \Bds decays to multibody charmless final states is also challenging. As for the two body decays the low branching fractions and the plethora of different final states makes very difficult to efficiently select the signals. Ror \lhcb, the analyses are further complicated by the low acceptance of the detector with respect to the long lifetime of the \KS. Recent results from \lhcb includes the measurement of the branching fractions of the various \decay{\Bds}{\KS\hp\hadron^{\prime -}} decays~\cite{Aaij:2017zpx} and the untagged time-integrated Dalitz analysis of the \decay{\Bd}{\KS\pip\pim} decay~\cite{Aaij:2017ngy} both performed with the full Run1 sample. The branching fractions are measured relatively to the \decay{\Bd}{\KS\pip\pim} and report the observation of all the \Bd and \Bs modes, except for the \decay{\Bs}{\KS\Kp\Km} for which only an upper limit is determined. The Dalitz analysis of the \decay{\Bd}{\KS\pip\pim } is the first amplitude analysis of a any \decay{\Bds}{\KS\hp\hadron^{\prime -}} mode at a hadronic machine. It  is described using the isobar model, including several intermediate resonances plus a non-resonant component. The \KS\pim S-wave is described using the EFKLLM model \cite{ElBennich:2009da}. The \CP-averaged fit fractions of all the components are determined as well as the \CP asymmetries of the flavour-specific final states. A very large \CP asymmetry is observed for the \decay{\Bd}{\Kstarm\pip} decay, corresponding to 
\begin{equation}
A_{\CP}(\decay{\Bd}{\Kstarm\pip}) = -0.308 \pm 0.062 \ ,
\end{equation}
which is the first observation of \CP violation in this channel with a significance of more than 6 standard deviations. The very large \CP asymmetry is visible in Figure~\ref{fig:kspipiCPV}, where the projection on the Dalitz-plot variables  $m^2_{\KS\pip}$ and (right) $m^2_{\KS\pim}$ are shown. The full potential of these decays can only be exploited with flavour-tagged time-dependent Dalitz analyses, that are very challenging and require much larger statistics. Thanks to the much better performances of flavour tagging, \belletwo will be able to remain competitive for the time-dependent measurement of the \Bd modes.

\subsection{\boldmath{\CP} violation in beauty baryons at LHCb}

In contrast to meson decays, \CP violation in the baryon sector has never been observed. In this quest, the study of beauty-baryon decays, and in particular those without charm quark in the final state, provides excellent opportunities. In fact, these decays might have large \CP asymmetries, as the tree-level and loop-level transitions are comparable in magnitude which allows for large interference effects. Thanks to the large production cross-section of beauty baryons in \proton\proton collisions at the \lhc, the \lhcb experiment is the only active experiment capable of expanding our knowledge
in this sector, as these decays are not accessible at the \ep\en KEK collider. The only measurement of \CP violation in \bquark-baryon decays previous to \lhcb was the measurement of the direct \CP asymmetries of \decay{\Lb}{\proton\Km} and \decay{\Lb}{\proton\pim} decays performed by \cdf~\cite{Aaltonen:2014vra} with precision of $\order(0.1)$. Using the full Run1 data sample, corresponding to integrated luminosities of 1\invfb and 2\invfb of \proton\proton collisions collected as \sqs=7\tev and \sqs=8\tev, respectively, \lhcb measured the same quantities with uncertainties reduced by approximately a factor 4~\cite{Aaij:2018tlk}
\begin{eqnarray*}
	A_{\CP}(\decay{\Lb}{\proton\Km}) & = & -0.020 \pm 0.013 \pm 0.019,\\
	A_{\CP}(\decay{\Lb}{\proton\pim}) & = & -0.035 \pm 0.017 \pm 0.020,
\end{eqnarray*}
where the first uncertainty is statistics and the second systematics. A measurement of the difference between the two \CP asymmetries is also performed, in which the main systematic uncertainties, being the asymmetry of detection efficiency between \proton and \antiproton, and the asymmetry between the production cross-section of \Lb and \Lbbar, cancel:
\begin{equation*}
A_{\CP}(\decay{\Lb}{\proton\Km})-A_{\CP}(\decay{\Lb}{\proton\pim}) = 0.014 \pm 0.022 \pm 0.013.
\end{equation*}

In addition, the \lhcb collaboration also studies $b$-hadron decays with multibody final states. Recently evidence for \CP violation in the \decay{\Lb}{\proton\pim\pip\pim} decay~\cite{Aaij:2016cla} with a significance exceeding the 3 standard deviations was reported. The analysis is based on the full Run1 statistics, corresponding to integrated luminosities of 1\invfb and 2\invfb of \proton\proton collisions collected as \sqs=7\tev and \sqs=8\tev, respectively, and the experiment has the potential to establish a first observation of \CP violation in a baryon decays when including also the Run2 data. In the analysis, triple products of the momenta of the final-state particles are used to build observables that are sensitive to the violation of the $P$ and \CP symmetry \cite{Durieux:2015zwa,Durieux:2016nqr}. These observables have the advantage of being almost insensitive to spurious asymmetries, like the detection asymmetry of charge-conjugate final states or the imbalance between the production cross sections between \Lb and \Lbbar baryons in \proton\proton collisions. The validity of this consideration is nevertheless tested by performing the analysis on the control mode \decay{\Lb}{\Lc\pim} in which \CP violation is expected to be negligible. The search for $P$ and \CP violation is also performed in different regions of the phase space of the decay. Two splitting scheme are used: one designated to isolate regions according to the dominant intermediate resonances and one designated to separate the sample in bins of the angle between the $\proton-\pim$ and the $\pip-\pim$ decay planes, as suggested in Refs.~\cite{Durieux:2015zwa,Durieux:2016nqr}. Also, the \decay{\Lb}{\proton\pim\Kp\Km} decay was studied by \lhcb, finding no evidence of $P$ and \CP violation ~\cite{Aaij:2016cla}. More recently the same technique has been applied to \decay{\Lb}{\proton\Km\pip\pim}, \decay{\Lb}{\proton\Kp\Km\Kp} and \decay{\Xibz}{\proton\Km\Kp\pim} decays, reconstructed using the Run1 data~\cite{Aaij:2018lsx}. Again, no evidence of $P$ or \CP violation is observed for these decay modes. Moreover, inspecting the phase space of the \proton\Km\pip\pim and \proton\Km\Kp\Km final states the decay \decay{\Lb}{\proton\Km\chiczero} is observed for the first time. The branching ratios of several \Lb and \Xibz decays to \proton3\Ph final states (with $h$ being charged kaons or pions) have also been measured relatively to the branching ratios of the \decay{\Lb}{\Lc(\proton\Km\pip)\pim}, using the same data sample~\cite{Aaij:2017pgy}. Given that the sector of beauty baryon decays is almost unexplored, the many results expected in the near future will require a joined effort both on the experimental and on the theory side to provide reliable interpretation of the experimental results.

\section{Conclusion}

The search for new sources of \CP violation in \B decays remains an interesting but challenging enterprise. With the start of \belletwo and the continued analyses at \lhcb, many exciting new results will surface in the near future. Among these are determination of $\gamma$ from single experiments at the few degree level and an expected combined precision of an impressive 1\degrees. In addition, with \belletwo running also the decays with neutral particles (like photons and \piz) in the final states will be updated and new \CP asymmetries measured, which provide also important guidance to stimulate theoretical progress. Finally, especially in the three- and multibody \B decays exciting results are expected in the nearby future. Most importantly, the full exploitation of this incredible amount of data, requires continued efforts and synergies between theorists and experimentalists. Theoretically, there are still the power-corrections and the description of three-body decays that require further investigations and many interesting avenues can still be explored, especially when making use of the available data. We look forward to the many interesting new analysis and theoretical progress that will be presented at the next CKM workshop. 

\section*{Acknowledgements}
We would like to thank all the speakers for WG V at CKM 2018 for their contributions; Joachim Brod, Alberto Correa Dos Reis, Resmi P.~K., Evelina Mihova Gersabeck,  Patricia Magalhães, Ruben Jaarsma, Tobias Huber, Vincent Tisserand, Bhubanjyoti Bhattacharya, Eli Ben-Haim, Emilie Bertholet, Peicheng Lu, Gediminas Sarpis, Bilas Pal, Timothy Gershon, Davide Fazzini, Luiz Vale Silva and Marcella Bona. KKV is supported by the Deutsche Forschungsgemeinschaft (DFG) within research unit FOR 1873 (QFET).

\bibliographystyle{LHCb}
\setboolean{inbibliography}{true}
\bibliography{proceedings}

\end{document}